\documentclass[runningheads]{llncs}

\usepackage[T1]{fontenc}
\usepackage{acro}
\usepackage{adjustbox}
\usepackage{booktabs}
\usepackage{comment}
\usepackage{todonotes}
\usepackage{colortbl}

\usepackage{array}
\usepackage{ragged2e}
\usepackage{booktabs}   
\usepackage{colortbl}
\usepackage{xcolor}
\usepackage{url}

\newcolumntype{a}{>{\columncolor{gray!10!white}}c}
\newcolumntype{x}{>{\columncolor{green!10!white}}c}
\newcolumntype{y}{>{\columncolor{blue!10!white}}c}
\newcolumntype{z}{>{\columncolor{yellow!10!white}}c}
\newcolumntype{v}{>{\columncolor{red!10!white}}c}
\definecolor{OliveGreen}{rgb}{0,0.6,0}
\definecolor{ForestGreen}{RGB}{34,139,34}
\definecolor{myblue}{RGB}{37,165,203}
\definecolor{FAUblue}{rgb}{0.000, 0.2196, 0.3961}
\definecolor{myred}{RGB}{175,32,67}

\colorlet{backgroundcol}{cyan!10!white}

\usepackage{array}
\newcolumntype{L}[1]{>{\raggedright\arraybackslash}p{#1}}

\usepackage{graphicx}
\usepackage{listings}
\usepackage[hidelinks]{hyperref}

\usepackage{float}

\definecolor{codegreen}{rgb}{0,0.6,0}
\definecolor{codegray}{rgb}{0.5,0.5,0.5}
\definecolor{codepurple}{rgb}{0.58,0.5,0.82}
\definecolor{backcolour}{rgb}{0.95,0.95,0.92}

\lstdefinestyle{mystyle}{
    backgroundcolor=\color{white},   
    commentstyle=\color{codegreen},
    keywordstyle=\color{magenta}\bfseries,
    moredelim=[is][\color{magenta}\bfseries]{@}{@},
    numberstyle=\tiny\color{codegray},
    stringstyle=\color{codepurple},
    basicstyle=\C\footnotesize,
    frame           = tb,         
    framerule       = 0.6pt,      
    rulecolor       = \color{black},
    framesep        = 0.4em,      
    xleftmargin     = 2em,
    framexleftmargin= 2em,
    breakatwhitespace=false,         
    breaklines=true,                 
    captionpos=b,                    
    keepspaces=true,                 
    numbers=left,                    
    numbersep=5pt,                  
    showspaces=false,                
    showstringspaces=false,
    showtabs=false,                  
    tabsize=2
}

\usepackage{amsmath,amssymb,amsfonts}

\usepackage{wrapfig}
\usepackage{graphicx}
\usepackage{soul}

\usepackage{multirow}

\begin{document}


\title{Computational Methods and GPU Acceleration in Plasma Physics: A Empirical Analysis of arXiv Publications and Research Trends}

\titlerunning{Computational Methods and GPU Acceleration in Plasma Physics}



\author{Jeremy J. Williams\inst{1}\and
Anders Brostr{ö}m\inst{1} \and
Stefano Markidis\inst{1} }
\authorrunning{Jeremy J. Williams et al.}
%
\institute{KTH Royal Institute of Technology, Stockholm, Sweden}

\maketitle


\begin{abstract}
Computational plasma physics increasingly relies on high-performance computing (HPC) methods, including particle-in-cell (PIC), gyrokinetic, and magnetohydrodynamic (MHD) simulations, yet field-wide evidence on how method choice, team size, and GPU adoption shape research outputs remains limited. We analyze 5,522 computational plasma physics papers published on arXiv between 2010 and 2025 using large-scale text mining, employing abstract length as a proxy for methodological and algorithmic complexity. Using ordinary least squares (OLS), tobit, and logistic regression models, we examine how abstract length and GPU mentions vary with computational method, number of authors, and publication year. Controlling for collaboration size and temporal trends, results show that MHD studies have longer abstracts than PIC and gyrokinetic papers, indicating more extensive methodological and physical exposition. Abstract length increases modestly with team size, while temporal effects suggest gradual changes in abstract conciseness over time. At the same time, PIC and gyrokinetic methods have grown substantially in relative prevalence over the past decade and are strongly linked to GPU adoption, reflecting their higher computational intensity and suitability for accelerator-based architectures. Together, these findings highlight a decoupling between methodological verbosity and method prevalence, offering a new bibliometric perspective on the evolution of HPC-driven plasma physics research.


\end{abstract}


\section{Introduction}

Plasma physics research is increasingly driven by large-scale computational simulations. Techniques such as particle-in-cell (PIC), gyrokinetic, and magnetohydrodynamic (MHD) models are central to understanding plasma dynamics across laboratory and astrophysical settings. The choice of numerical method, coupled with the adoption of GPUs and other HPC resources, reflects both the underlying physics problems and computational constraints.

While detailed case studies exist for specific codes (e.g., BIT1~\cite{williams2025accelerating,williams2023leveraging}, GENE/GENEX ~\cite{trilaksono2025openacc,trilaksono2024characterizing}, JOREK~\cite{williams2024understanding} and Vlasiator~\cite{coti2024integration}), there is limited evidence on field-wide publication patterns that link computational methods, team composition, and algorithmic complexity. Understanding these patterns can illuminate trends in method sophistication, collaboration, and HPC adoption.

This work introduces a bibliometric and econometric approach to quantify computational complexity across plasma physics research. We use abstract length as a proxy for methodological and algorithmic complexity, leveraging a curated dataset of computational plasma physics papers from arXiv (2010–-2025). Our central research question is: Do abstract lengths differ systematically across computational methods in plasma physics, and how are they influenced by team size and temporal trends? The main contributions of this work can be summarized as follows:

\begin{itemize}
\item We curated and analyzed a dataset of 5,522 computational plasma physics research articles from the arXiv repository (2010–-2025).
\item We developed a systematic filtering pipeline to identify papers focused on numerical methods and HPC-based simulations, including PIC, gyrokinetic, and MHD approaches.
\item We quantified methodological and algorithmic complexity using abstract length as a proxy, and examined the influence of team size, computational method, and publication year.
\item We applied ordinary least squares (OLS) regression, complemented by tobit and logistic models, to rigorously assess field-wide patterns in computational methods and GPU adoption.
\item We visualized trends over time using heatmaps, line charts, and area plots, highlighting method prevalence, abstract length distributions, and GPU adoption across the computational plasma physics literature.
\end{itemize}

\section{Methodology \& Experimental Setup}

In this work, we aim to explore computational methods and GPU acceleration in plasma physics by analyzing publication patterns across the field. Our goal is to quantify methodological complexity using abstract length as a proxy, and to examine how abstract verbosity is influenced by computational method, team size, and publication year.

\subsection{Data Collection \& Filtering}

The input data for this work consists of research papers obtained from the arXiv repository using the \texttt{arxiv} Python API, which allows systematic querying of metadata from the e-print service (\url{https://arxiv.org}). We focused on the \texttt{physics.plasm-ph} category (Plasma Physics), with key metadata extracted for each paper including:

\begin{itemize}
    \item \textbf{Title} of the paper
    \item \textbf{Abstract} summarizing the research
    \item \textbf{Authors} of the publication
    \item \textbf{Number of authors}
    \item \textbf{Publication date} (year and month)
    \item \textbf{URL} of the arXiv entry
\end{itemize}

A filtering pipeline was applied to collect papers relevant to \emph{computational plasma physics}. The criteria were as follows:

\begin{itemize}
    \item \textbf{Included if:} the title or abstract contained keywords such as \emph{PIC, particle-in-cell, gyrokinetic, gyrofluid, Vlasov, Monte Carlo, Poisson solver, numerical solver, parallelization, GPU, code architecture}, etc.
    \item \textbf{Excluded if:} the text primarily referenced experimental setups or diagnostics, such as \emph{probe, instrumentation, spacecraft, spectroscopy, antenna, satellite}, etc.
    \item \textbf{Retained if:} the abstract included explicit computational context, such as \emph{solver, numerical, simulate, implement, compute}, etc.
\end{itemize}

\subsection{Data Description \& Overview}
The resulting dataset represents a curated collection of computational plasma physics articles extracted and filtered from arXiv. Articles published between \texttt{01/2010} (January 2010) and \texttt{12/2025} (December 2025) were considered. The data collection process queried all months in this interval (192 months total), yielding:

\begin{itemize}
    \item \textbf{16,590} Plasma Physics (\texttt{physics.plasm-ph}) papers retrieved from arXiv
    \item \textbf{5,522} papers retained after computational-method filtering
\end{itemize}

\noindent The final enriched dataset was saved and named as \texttt{"plasma\allowbreak\_computational\allowbreak\_methods\allowbreak\_regression.csv"}. In addition to bibliographic metadata, each record includes derived variables for subsequent statistical analysis:

\begin{itemize}
    \item \textbf{Abstract length} (number of words)
    \item \textbf{Censored abstract length} for regression modeling (capped at 500 words)
    \item \textbf{Method category} (e.g., PIC, gyrokinetic, MHD, other)
    \item \textbf{Binary GPU indicator} based on whether GPUs were mentioned (yes = 1, no = 0)
\end{itemize}

Table~\ref{table:top_5_papers} shows the top five articles retained after applying the filtering. 


\begin{table}[!ht]
\vspace{0cm} 
\centering
\resizebox{\textwidth}{!}{%
\renewcommand{\arraystretch}{1.15}
\setlength{\tabcolsep}{4pt}
\begin{tabular}{|c|
                L{3.2cm}|
                L{4.8cm}|
                L{3.6cm}|
                c|c|c|c|
                L{3.4cm}|
                c|c|c|c|}
\hline
\rowcolor{lightgray}
\textbf{id} &
\textbf{title} &
\textbf{abstract} &
\textbf{authors} &
\textbf{n\_authors} &
\textbf{published} &
\textbf{year} &
\textbf{month} &
\textbf{url} &
\textbf{abstract\_len} &
\textbf{mentions\_gpu} &
\textbf{method\_category} &
\textbf{abstract\_len\_cens} \\
\hline

\cellcolor{gray!10} 0 &
Semianalytical treatment of current density of particles injected by a monoenergetic source &
This paper extends the semianalytical treatment of fast ion current density to include time dependence and velocity diffusion using solutions of the Boltzmann equation with a complete Coulomb collision term... &
P. R. Goncharov; B. V. Kuteev; V. Yu. Sergeev; T. Ozaki; S. Sudo &
5 &
2010-04-28 &
2010 &
4 &
\url{http://arxiv.org/abs/1004.5103v2} &
106 &
0 &
Other &
106 \\
\hline

\cellcolor{gray!10} 1 &
The lower hybrid wave cutoff: A case study in eikonal methods &
Eikonal or ray-tracing methods are used to estimate radio-frequency field propagation in plasmas. This work compares standard eikonal approaches with wave-packet based dynamics... &
A. S. Richardson; P. T. Bonoli; J. C. Wright &
3 &
2010-04-28 &
2010 &
4 &
\url{http://arxiv.org/abs/1004.4943v1} &
136 &
0 &
Other &
136 \\
\hline

\cellcolor{gray!10} 2 &
Analysis of plasma instabilities and verification of the BOUT code for the Large Plasma Device &
Linear instabilities in the Large Plasma Device are studied using analytic theory and numerical solutions of collisional plasma fluid equations with the 3D BOUT code... &
P. Popovich; M. V. Umansky; T. A. Carter; B. Friedman &
4 &
2010-04-26 &
2010 &
4 &
\url{http://arxiv.org/abs/1004.4674v4} &
144 &
0 &
Other &
144 \\
\hline

\cellcolor{gray!10} 3 &
Invariants, symmetries and scaling of Rayleigh--Taylor turbulent mixing &
Scaling laws, invariants, and spectral properties of unsteady turbulent mixing induced by the Rayleigh--Taylor instability are analyzed using extensions of Kolmogorov theory... &
Snezhana I. Abarzhi &
1 &
2010-04-26 &
2010 &
4 &
\url{http://arxiv.org/abs/1004.4670v1} &
81 &
0 &
PIC &
81 \\
\hline

\cellcolor{gray!10} 4 &
Strong magnetohydrodynamic turbulence with cross helicity &
High-resolution numerical simulations of incompressible MHD turbulence with and without cross helicity are presented, showing scale-invariant inertial-range behavior... &
Jean Carlos Perez; Stanislav Boldyrev &
2 &
2010-04-21 &
2010 &
4 &
\url{http://arxiv.org/abs/1004.3798v1} &
148 &
0 &
MHD &
148 \\
\hline
\end{tabular}
} 

\vspace{0.4em}
\caption{Top five entries from the \texttt{"plasma\_computational\_methods\_regression.csv"} dataset, showing all metadata and derived variables used for filtering and regression analysis. The \texttt{mentions\_gpu} column is a binary indicator based on keyword detection.}
\label{table:top_5_papers}
\vspace{-0.6cm} 
\end{table}

\subsubsection{Data Density Heatmap:}
\begin{figure}[!ht]
\vspace{0cm} 
    \centering
    \includegraphics[width=0.95\linewidth]{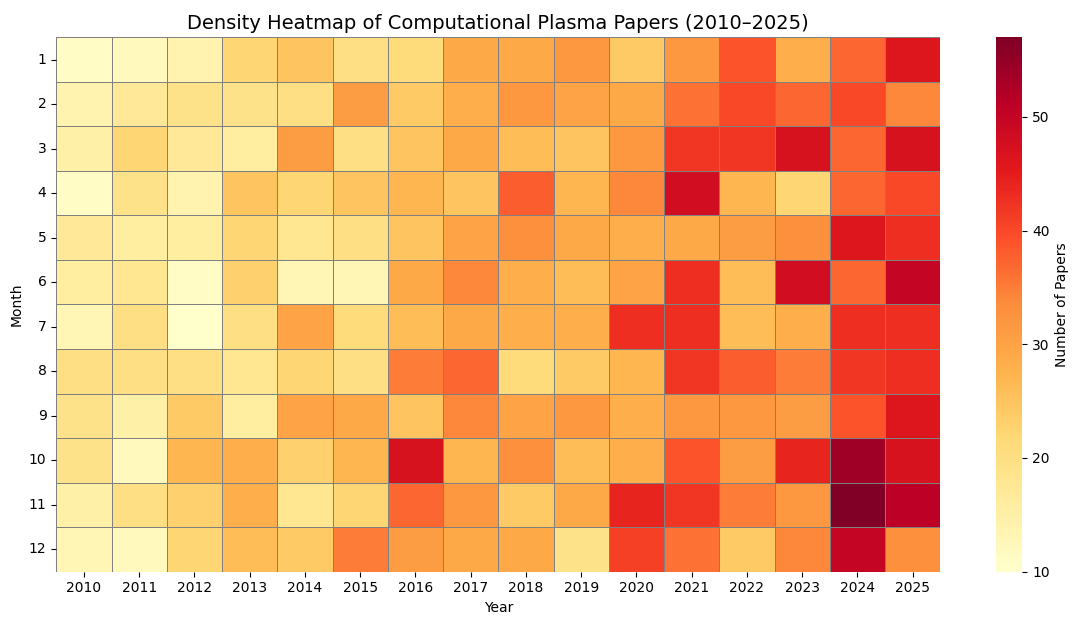}
    \caption{Density Heatmap of computational plasma research papers published from 2010 to 2025. Darker colors indicate months with higher numbers of papers, highlighting periods of increased research activity over time.}
    \label{fig:density_heatmap_papers}
    \vspace{-0.6cm} 
\end{figure}

The final dataset can be visualized to understand the distribution of computational plasma papers over time. A density heatmap was generated, where each cell corresponds to the number of filtered papers published in a given month/year. Fig.~\ref{fig:density_heatmap_papers} shows the monthly publication counts of computational plasma papers from 2010 to 2025. The X-axis represents years, the Y-axis represents months, and darker colors indicate months with higher numbers of papers, highlighting trends and periods of increased research activity, with greater publication density observed after 2015.

\subsection{Regression Framework}

\subsubsection{OLS Regression.}
To investigate whether abstract lengths differ systematically across computational methods and how they relate to team size, publication year, and GPU adoption, we employ OLS regression, complemented by Tobit and logistic models to ensure robustness. The OLS model is specified as:

\begin{equation}
\text{abstract\_len}_i = \beta_0 + \beta_1 ,\text{n\_authors}_i + \beta_2 ,\text{year}_i + \sum_m \beta_{3m} ,\text{method}_{im} + \varepsilon_i
\label{eq:ols_abstract_len}
\end{equation}

\noindent where:

\begin{equation*}
\begin{cases}
\text{abstract\_len}_i &: \text{number of words in the abstract of paper } i,\\
\text{n\_authors}_i &: \text{number of authors on paper } i,\\
\text{year}_i &: \text{publication year of paper } i,\\
\text{method}_{im} &: \text{dummy variable } m , \text{(PIC, gyrokinetic, MHD (baseline))},\\
\beta_0 &: \text{intercept (expected abstract length for MHD papers)},\\
\beta_1 &: \text{effect of team size on abstract length},\\
\beta_2 &: \text{temporal trend in abstract verbosity},\\
\beta_{3m} &: \text{additional words for method } m \text{ relative to MHD},\\
\varepsilon_i &: \text{error term capturing unobserved factors.}
\end{cases}
\end{equation*}




\subsubsection{Tobit Regression.} 
Since some abstracts are censored at an upper bound (e.g., 500 words), we also employ a Tobit regression to account for truncation:

\begin{align}
\text{abstract\_len\_cens}_i^* &= \beta_0 + \beta_1 \,\text{n\_authors}_i + \beta_2 \,\text{year}_i + \sum_m \beta_{3m} \,\text{method}_{im} + \varepsilon_i, \\
\text{abstract\_len\_cens}_i &=
\begin{cases}
\text{abstract\_len\_cens}_i^* & \text{if } \text{abstract\_len\_cens}_i^* \le 500,\\
500 & \text{if } \text{abstract\_len\_cens}_i^* > 500
\end{cases}
\label{eq:tobit_abstract_len}
\end{align}

\noindent where:

\begin{equation*}
\begin{cases}
\text{abstract\_len\_cens}_i^* &: \text{latent (true) abstract length of paper } i,\\
\text{abstract\_len\_cens}_i &: \text{observed abstract length, censored at 500 words},\\
\beta_0 &: \text{intercept (expected abstract length for MHD papers)},\\
\beta_1 &: \text{effect of team size on abstract length},\\
\beta_2 &: \text{temporal trend in abstract length},\\
\beta_{3m} &: \text{additional words for method } m \text{ relative to MHD},\\
\varepsilon_i &: \text{error term capturing unobserved factors.}
\end{cases}
\end{equation*}

\noindent where $\text{abstract\allowbreak\_len\allowbreak\_cens}_i^*$ is the latent (true) abstract length \& $\text{\allowbreak abstract\allowbreak\_len\allowbreak\_cens}_i$ is the observed length, censored at 500 words. Tobit regression estimates the effects of team size, year, and method while accounting for this upper limit.

\subsubsection{Logistic Regression.} 
Finally, to model GPU adoption as a binary outcome (whether a paper mentions GPU acceleration), we apply a logistic regression:
{\footnotesize
\begin{equation}
\text{logit} , P(\text{mentions\_GPU}_i = 1) = \alpha_0 + \alpha_1 ,\text{n\_authors}_i + \alpha_2 ,\text{year}_i + \sum_m \alpha_{3m} ,\text{method}_{im}
\label{eq:logit_gpu}
\end{equation}
}
\noindent where:

\begin{equation*}
\begin{cases}
\text{mentions\_GPU}_i = 1 & \text{if GPU acceleration is mentioned in paper } i,\\
\text{mentions\_GPU}_i = 0 & \text{otherwise},\\
\alpha_0 &: \text{baseline log-odds of GPU mentioned (MHD papers)},\\
\alpha_1 &: \text{effect of team size on log-odds of mentioning GPU},\\
\alpha_2 &: \text{temporal trend effect on GPU mentions},\\
\alpha_{3m} &: \text{effect of computational method } m \text{ relative to MHD}.
\end{cases}
\end{equation*}


\subsection{Hardware \& Software Environment}
All experiments and analyses were conducted using \texttt{Google Colab}, leveraging its convenient cloud-based environment. We used \texttt{Python 3} as the programming language, with the runtime type set to either the \texttt{Google Compute Engine backend} (\texttt{CPU}) or (\texttt{A100} / \texttt{L4} / \texttt{T4 GPU}) depending on the computational requirements. The \texttt{CPU} runtime was employed for lighter preprocessing and small-scale tasks, whereas the \texttt{A100} / \texttt{L4} / \texttt{T4 GPU} accelerator enabled faster execution of computationally intensive operations, such as generating embeddings, clustering, and creating visualizations. This setup allowed us to efficiently process the dataset and generate plots for analysis while taking advantage of the \texttt{GPU} for parallelized computation when needed.

\section{ArXiv Article Analysis and Results} \label{sec:results}
In this work, we analyze computational plasma physics research articles collected from the arXiv repository to investigate how computational methods, team size, and GPU adoption influence research outputs. Using a curated dataset of 5,522 papers from 2010 to 2025, we focus on quantifying methodological and algorithmic complexity through abstract length as a proxy.

\begin{figure}[!ht]
    \centering
    \includegraphics[width=0.75\linewidth]{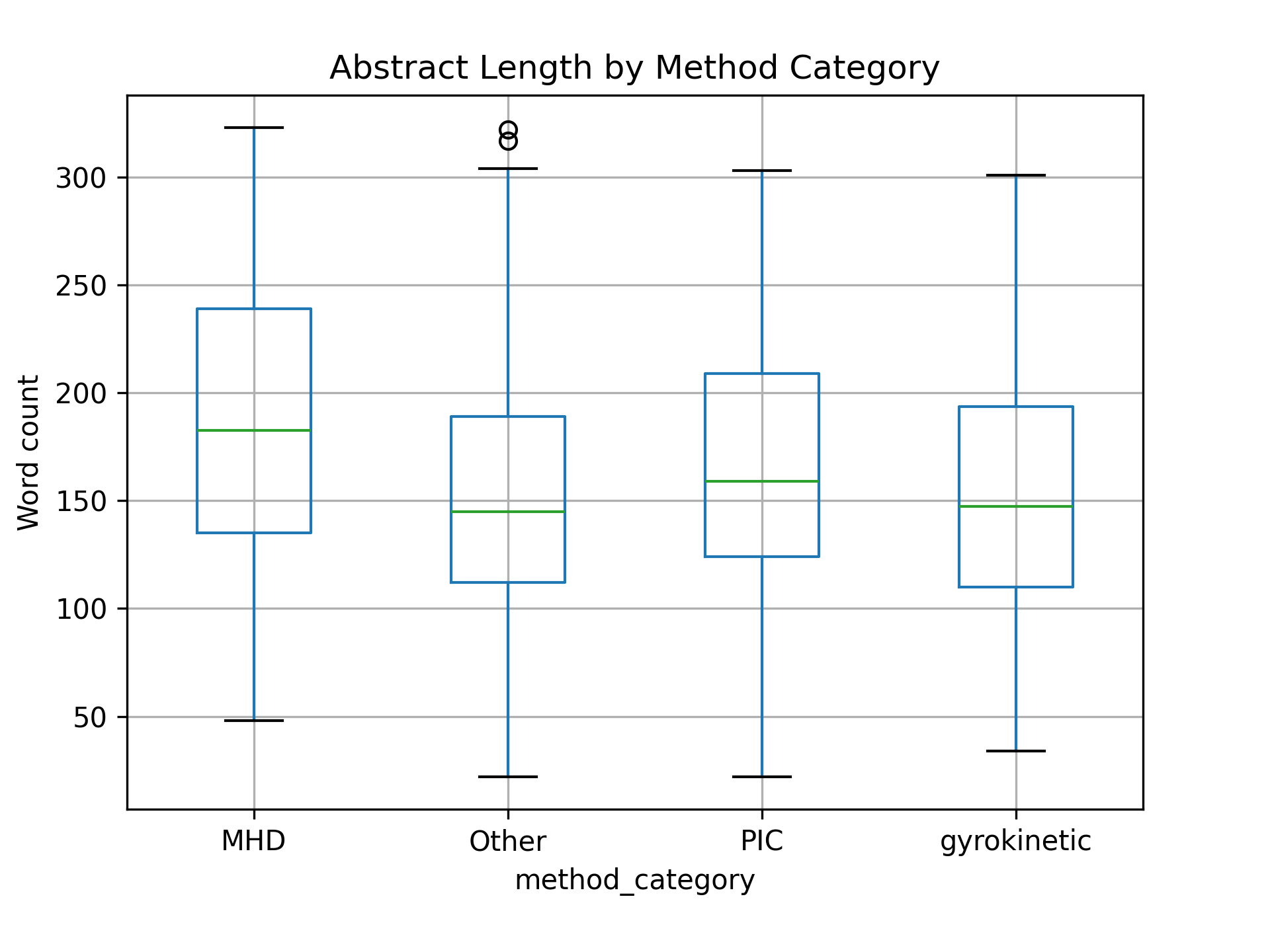}
    \caption{Boxplot of abstract lengths by computational method category. MHD papers generally have longer abstracts than PIC and gyrokinetic papers.}
    \label{fig:abstract_length_boxplot}
\end{figure}

We begin by examining the distribution of abstracts by computational method. Fig.~\ref{fig:abstract_length_boxplot} shows that MHD studies tend to have longer abstracts than PIC and gyrokinetic papers, suggesting more detailed methodological descriptions. Abstract length also correlates positively with the number of authors, indicating that larger collaborative teams provide more comprehensive abstracts.

\begin{figure}[!ht]
    \centering
    \includegraphics[width=0.75\linewidth]{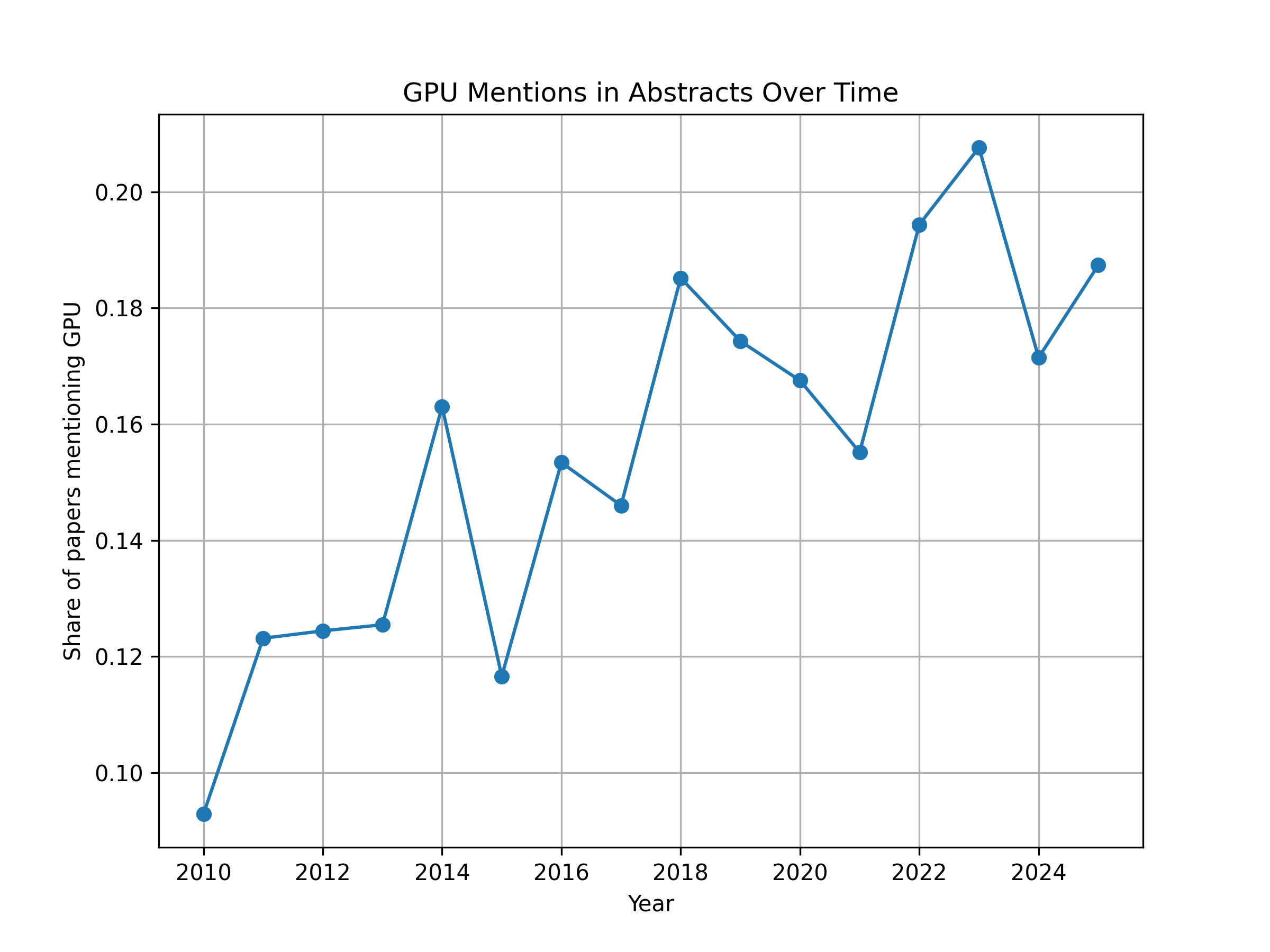}
    \caption{Share of papers mentioning GPUs in abstracts over time (2010–-2025), showing the increasing adoption of GPU-accelerated computing.}
    \label{fig:gpu_adoption_trend}
\end{figure}

Next, we analyze trends over time. The monthly and yearly publication counts (Fig.~\ref{fig:density_heatmap_papers}) highlight growth in computational plasma physics research, with a notable increase in papers after 2015. GPU mentions in abstracts have risen steadily, reflecting broader adoption of high-performance computing resources across the field (Fig.~\ref{fig:gpu_adoption_trend}).

\begin{figure}[!ht]
\vspace{0cm} 
    \centering
    \includegraphics[width=0.85\linewidth]{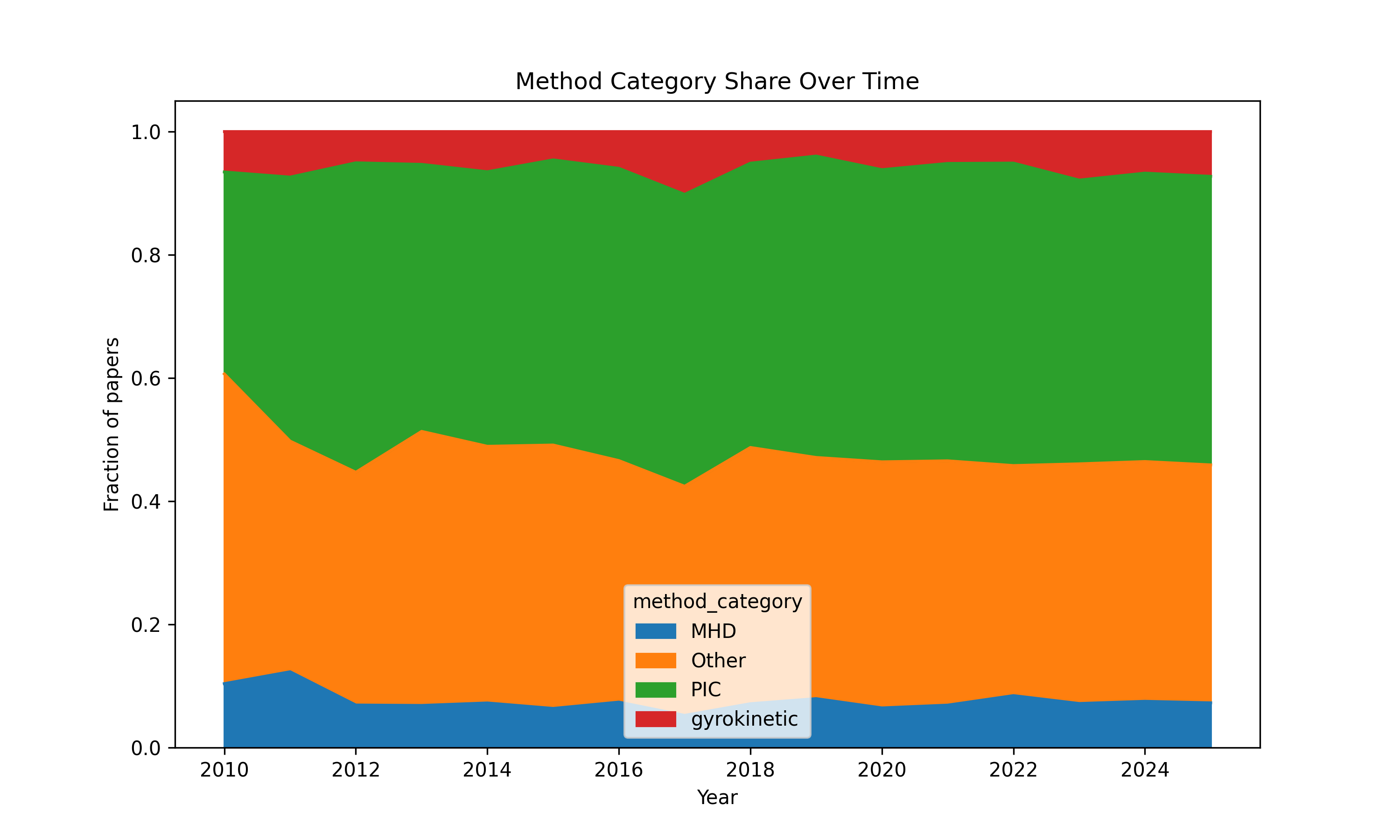}
    \caption{Relative share of computational method categories over time from 2010 to 2025, showing increased prevalence of PIC and gyrokinetic approaches compared to MHD.}
    \label{fig:method_share_trend}
\end{figure}

Finally, we examined changes in the relative prevalence of computational methods over time. Fig.~\ref{fig:method_share_trend} shows that, although MHD studies tend to have longer abstracts than PIC and gyrokinetic methods, the share of PIC and gyrokinetic papers has increased over the past decade, while the relative share of MHD papers has declined. This shift is consistent with the rise in GPU adoption, as PIC and gyrokinetic methods are often more computationally intensive and benefit significantly from accelerator-based architectures. 

\subsection{Regression Results.}
To formally test the patterns observed in abstract lengths across computational methods, team size, and publication year, we apply OLS, tobit, and logistic regression models to ensure robustness in our findings.

\subsubsection{OLS Model.}
We use Eq.~\ref{eq:ols_abstract_len} to model abstract length as a function of publication year, number of authors, and computational method category (baseline = MHD). 

\begin{figure}[!ht]
\vspace{0cm} 
    \begin{center}
        \includegraphics[width=\linewidth]{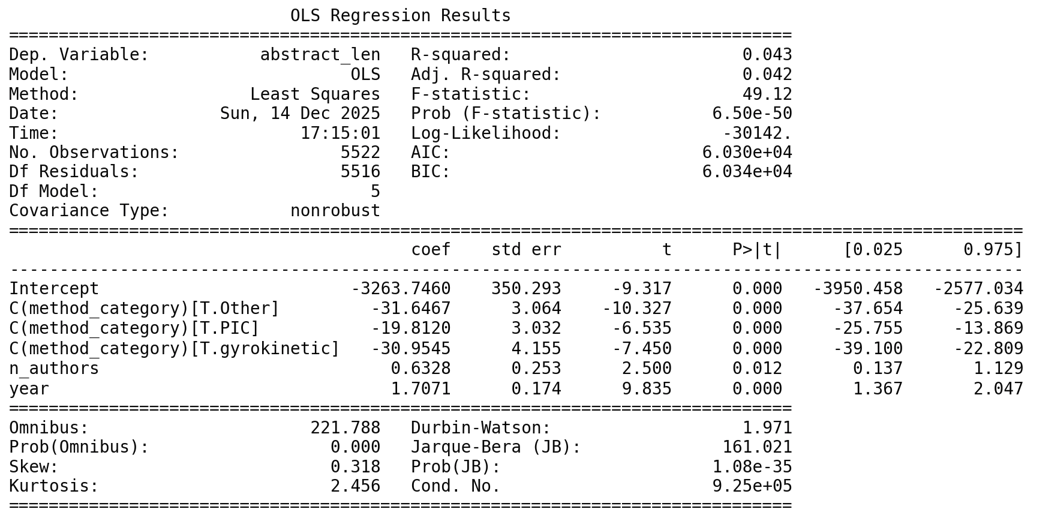}
        \caption{OLS Regression Results for Eq.~\ref{eq:ols_abstract_len}.} 
        \label{OLS_Model_Summary-Final}
    \end{center}
\end{figure}

As seen in Fig.~\ref{OLS_Model_Summary-Final}, regression results show that MHD methods are associated with significantly longer abstracts compared to both PIC and gyrokinetic methods, controlling for team size and publication year. Specifically, the coefficients for PIC and gyrokinetic methods are negative, indicating that these methods contribute fewer words in abstracts relative to MHD. The number of authors also has a positive effect, with each additional author contributing to a modest increase in abstract length. Additionally, there is a temporal trend, with abstract lengths increasing over time. The $R^2$ value for the model is $0.043$, indicating that the predictors explain a small portion of the variation in abstract length, but the model is still significant ($p$-value $< 0.0001$). The coefficients for method categories indicate that PIC and gyrokinetic methods contribute fewer words compared to MHD, with the strongest effect observed for gyrokinetic methods.

\newpage
\subsubsection{Tobit Model.}
Since some abstracts are censored at an upper bound (e.g., 500 words), we complement the OLS analysis with a tobit regression (Eq.~\ref{eq:tobit_abstract_len}) to account for truncation. 

\begin{figure}[!ht]
\vspace{0cm} 
    \begin{center}
        \includegraphics[width=0.95\linewidth]{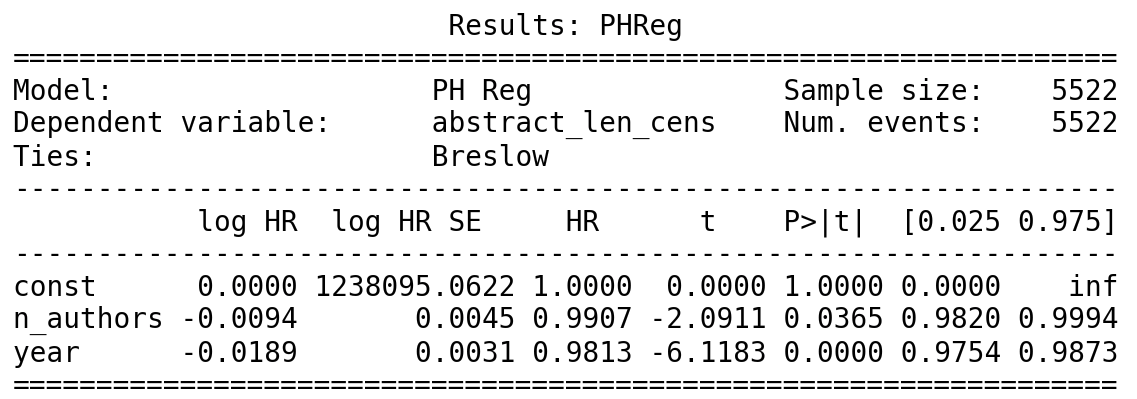}
        \caption{Tobit Regression Results for Eq.~\ref{eq:tobit_abstract_len}} 
        \label{Tobit_Model_Summary-Final}
    \end{center}
\end{figure}

Fig.~\ref{Tobit_Model_Summary-Final} results confirm the findings from the OLS regression, showing that the number of authors and the computational method have significant effects on abstract length. The latent (true) abstract length for papers that are censored at 500 words indicates that MHD methods result in longer abstracts compared to PIC and gyrokinetic methods. The effect of the number of authors is negative on the hazard ratio ($HR = 0.9907$), meaning that for each additional author, the true (latent) abstract length decreases marginally, though this effect is statistically significant ($p-value = 0.0365$).

Additionally, the temporal trend remains consistent, with abstract lengths becoming shorter over time, as indicated by the coefficient for the year variable ($log HR = -0.0189$, $p-value < 0.0001$), which suggests that over time, abstracts are becoming more concise. 

\subsubsection{Logistic Model.}
Next, we model GPU adoption as a binary outcome (whether a paper mentions GPU acceleration, general GPU usage, GPU computing, or the need for hardware accelerators due to emerging technological challenges), we apply logistic regression (Eq.~\ref{eq:logit_gpu}).

\begin{figure}[!ht]
\vspace{0cm} 
    \begin{center}
        \includegraphics[width=\linewidth]{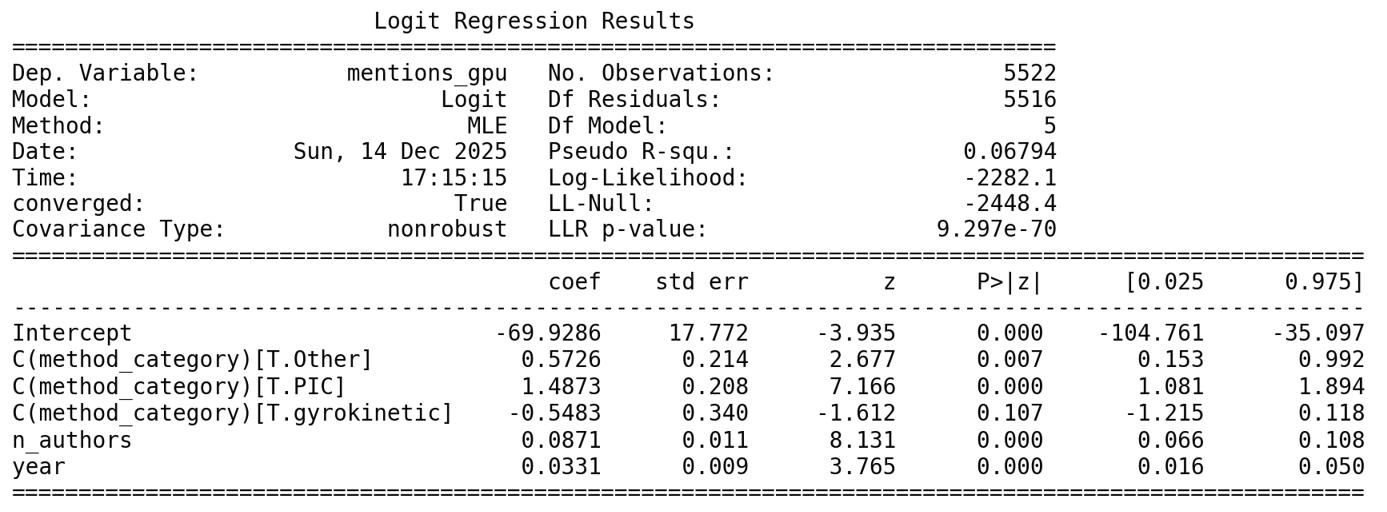}
        \caption{Logistic Regression Results for Eq.~\ref{eq:logit_gpu}} 
        \label{Logistic_Model_Summary-Final}
    \end{center}
\end{figure}

Fig.~\ref{Logistic_Model_Summary-Final} results show that the number of authors and the computational method significantly influence the likelihood of mentioning GPU acceleration. Specifically, PIC methods are strongly associated with mentioning GPUs, with a coefficient of $1.4873$ and a highly significant p-value of $p < 0.0001$. In contrast, gyrokinetic methods are not significantly associated with GPU mentions, as indicated by a p-value of $0.107$. The number of authors also has a positive effect on the likelihood of mentioning GPUs, with each additional author increasing the likelihood by $0.0871$ (p-value $< 0.0001$). Additionally, there is a temporal trend with an increase in GPU mentions over time, as indicated by the coefficient for the year variable ($0.0331$, $p-value < 0.0001$).

\section{Related Work}
Empirical studies examining computational methods in plasma physics at a field-wide level are limited. Most prior research has focused on case studies of specific simulation code performance~\cite{faj2023mpi,trilaksono2024characterizing,williams2023leveraging,williams2024characterizing,williams2024understandingV2}, optimizations~\cite{williams2023leveraging,williams2024optimizing,williams2025accelerating,williams2026high,williams2026multi}, I/O monitoring~\cite{williams2024enabling,williams2024understanding}, parallelization strategies~\cite{trilaksono2025openacc}, data streaming~\cite{williams2026integrating}, HPC workloads~\cite{medeiros2025arc}, enabling efficient vectorization~\cite{chronaki2026enabling} and hybrid quantum-classical benchmarking~\cite{hegde2025hybrid}, without systematically analyzing trends across the broader literature. This gap motivated our bibliometric and econometric approach, in which we analyzed 5,522 arXiv publications from 2010 to 2025 to investigate how computational method choice, team size, and GPU adoption influence research outputs. Our methodology draws on well-established statistical models for censored and limited data. Tobit models, for example, are widely used to handle dependent variables constrained by an upper or lower bound, ensuring unbiased coefficient estimates when ordinary linear regression is insufficient~\cite{amemiya1984tobit,amore2021tobit}. Similarly, logistic and probit models are designed for binary or categorical outcomes, allowing robust inference when standard linear assumptions do not hold~\cite{carroll1993robustness,guneri2020dependent}. The work by Nassimbeni~\cite{nassimbeni2001technology} also employs a logit-Tobit model to analyze how technological and innovation capacity influence export attitudes, showing how logistic regression can predict outcomes based on key factors, offering insights into empirical modeling approaches relevant to our study.


\section{Discussion \& Future Work}
Analyzing 5,522 computational plasma physics research articles from arXiv revealed systematic patterns in how research was communicated across different computational methods. By examining abstract lengths, team sizes, and temporal trends, we identified differences not apparent from simple publication counts or keyword searches. MHD studies had the longest abstracts, while PIC and gyrokinetic papers were generally shorter, reflecting differences in methodological and algorithmic complexity. Larger collaborative teams produced more comprehensive abstracts, suggesting that diverse expertise and collective effort contributed to more detailed reporting of research methods. Temporal trends were nuanced: while simple regression suggested a slight increase in abstract lengths over time, models accounting for censored abstracts indicated a shift toward more concise abstracts in recent years. Visualizations, including boxplots of abstract lengths and temporal trend plots, further illustrated these patterns and highlighted the steady growth of GPU adoption, particularly in PIC studies. Together, these analyses confirm that abstract lengths systematically differ across computational methods and are influenced by team size and temporal trends, demonstrating how methodological choices, collaborative scale, and temporal evolution shape how researchers communicate their work.

Building on these findings, future research could extend this analysis by incorporating full-text sections, such as methods and results, to capture deeper insights into computational complexity and scientific rigor. Integrating citation networks, co-authorship dynamics, and links to experimental or simulation datasets could reveal how knowledge propagates across the field. Additionally, leveraging large language models (LLMs) could enable semantic analysis of abstracts and full texts, uncovering hidden patterns, emerging trends, and potential interdisciplinary connections. These extensions would provide powerful tools to dynamically track scientific progress, guide future research directions, and foster collaboration across computational plasma physics and related domains.

\bibliographystyle{splncs04}
\bibliography{main} 

\end{document}